\documentclass[12pt,a4paper]{article} 
\usepackage{subeqn}
\usepackage{graphicx} 

\newcommand{\vett}[1]{\mathbf{#1}}
\newcommand{\vect}[1]{\vett #1}
\newcommand{\tdot}[1]{{\hskip2pt\ddot{\null}\hskip2.5pt%
    \dot{\null}\kern -5pt {#1}}}

\newcommand{\etal}{et al.}
\newcommand{\Name}[1]{\textrm{#1}, } 
\newcommand{\REVIEW}[4]{\textsl{#1} \textbf{#2}, #3 (#4)} 
\newcommand{\Book}[1]{\textsl{#1}, } 
\newcommand{\Editor}[1]{ed. \textrm{#1}, }
\newcommand{\Year}[1]{#1)} 
\newcommand{\Publ}[1]{(#1,} 
\newcommand{\Page}[1]{#1}

\title{Density limit in a first principles model\\
 of a magnetized plasma\\
in the Debye--H\"uckel approximation}

\author {A. Carati\thanks{Dipartimento di Matematica, Universit\`a
                          degli Studi di Milano, Milano, Italy } \and
         M. Zuin\thanks{Consorzio RFX, Associazione EURATOM-ENEA 
	                sulla Fusione, Padova, Italy} \and
	 A. Maiocchi\thanks{Dipartimento di Matematica, Universit\`a
                            degli Studi di Milano, Milano, Italy } \and
	 M. Marino\thanks{Dipartimento di Matematica, Universit\`a
                          degli Studi di Milano, Milano, Italy } \and
	 E. Martines\thanks{Consorzio RFX,   Associazione EURATOM-ENEA 
	                  sulla Fusione, Padova, Italy} \and 
	 L. Galgani\thanks{Dipartimento di Matematica, Universit\`a
                           degli Studi di Milano, Milano, Italy }
	 } 

\date{\today}

\begin{document}

\maketitle

\begin{abstract}
A crucial problem concerning a large variety of   
fusion devices is that   the confinement
due to an external magnetic field is lost above a critical  density,
 while a widely accepted
first principles explanation of such a fact is apparently lacking.
In the present paper,
making use of standard methods of statistical
mechanics in the Debye--H\"uckel approximation, we give indications
 that for a plasma there
exists a density threshold corresponding to a
transition from order to chaos, the ordered motions being those in
which  the confining
Lorentz force on a single electron  prevails over the diffusive effect of the
Coulomb forces. The    
density limit, which  is proportional to the square of the
magnetic field, turns out  to  fit  not too badly the
empirical data for plasma collapses in a large set of fusion devices. 
\end{abstract}

\section{Introduction} 
It is well known that in  most fusion devices 
a density limit for plasma confinement exists, while ``\emph{there is no
widely accepted  first principles model}'' for it (see the review  
\cite{greenwald}).
In the present paper, using the methods of statistical mechanics
in the Debye--H\"uckel approximation,
we show  that for a magnetized plasma  there exists  a transition from  order to
chaos at a critical  density. 
By ordered motions we just mean those for which the confining magnetic Lorentz
force acting on a single electron prevails against the sum of the  diffusive
 Coulomb forces of all the other particles, 
so that gyration  prevails over diffusion.

The  transition  occurs beyond a density limit which
is comparable to that observed for plasma collapses in a large set of
 fusion devices.  
Let us recall that, according to Alfv\'en \cite{alfven} (see also
page 382 of the first scientific  paper \cite{bohr} of Bohr), 
a  plasma can present a
diamagnetic behavior (and thus a confining pressure) only when it is
in an out--of--equilibrium state,
which thus can persist only in the presence of  sufficiently ordered
motions,  corresponding to the existence of a suitable adiabatic
invariant. Instead, in the presence of a strong
chaoticity, diamagnetism is quickly
lost (see \cite{chaos}). 
This is the reason why the transition from order to chaos discussed
here may be related to the collapses observed in fusion devices.

So the problem is that of comparing  the size of the
 confining Lorentz force acting on a single electron to that of the
diffusive force  due to the Coulomb interactions with all the other
particles. The estimate is
trivial for the confining Lorentz force, for
it is sufficient to estimate the velocity of the electrons, and this is
immediately obtained in terms of temperature. For what concerns
the size of the diffusive force  due to all the Coulomb interactions,
we  estimate it
using the standard methods of  statistical  mechanics, which compel one
to take into account the $n$--body collisions for all $n$, so that
the Debye length $\lambda_D$  then naturally enters into play (a comment
on the possibility of using the methods of kinetic theory instead of the general 
ones of statistical mechanics will be given later).

Now, rigorous  estimates of the sum of the Coulomb
forces on a single particle can be found in the literature 
 only  for systems of pure electrons. Indeed,  the Gibbs
distribution can be consistently dealt with  in such a case,
because the results do not depend too strongly  on the cutoffs that
have to be introduced  to manage the long range character of the
forces. Instead, in the presence of
the  neutralizing ions  one meets with  the problem that the
potential energy is not bounded from below. This requires introducing
short range cutoffs, and the results  now strongly depend  on them,
i.e., are model dependent (for a  general
 introduction to the problem see \cite{bogol},  \cite{meeron2},
\cite{meeron},  \cite{fisher} and the huge bibliography 
in \cite{review}). On the other hand, one expects that the
force due to the ions makes the system more chaotic, so that, in order
to give a model independent estimate of the threshold, in the present paper we limit
ourselves to considering the contribution of
the electrons, which should correspond to giving an upper bound to the threshold.  

We will show that this leads to a theoretical  electron density limit
$n_e$ given by 
 the law
\begin{equation}\label{sei}
n_e=  3\, \frac {B^2}{\mu_0 mc^2}\ ,
\end{equation}
where $B$ is the magnetic field, $\mu_0$ the vacuum permeability, $c$
the speed of light and $m$ the electron mass.
Using $n_e=1/a^3$ where $a$ is the mean interelectron 
distance, relation (\ref{sei}) can also be put 
in the  particularly expressive form
\begin{equation}\label{sette}
\frac 1{2\mu_0}B^2a^3= \frac 16\, mc^2\ ,
\end{equation}
according to which the   transition from order to chaos occurs when
the magnetic energy inside a  cell of volume $a^3=1/n_e$ just equals
(apart from a factor   $1/6$) the electron rest energy $mc^2$. 

We will show below  that the theoretical formula (\ref{sei}) fits not too 
badly the
phenomenological density limit for plasma collapses in a large set of
fusion devices, and this suggests
that the contributions of the ions, which we have here neglected,
should  be of the same order of magnitude as that of the
electrons. This conjecture is supported by a further result, which will
be published elsewhere (see \cite{noi} for the  case of a smeared out
background). Namely, one
can discuss the stability properties of the equilibrium solutions 
of a neutral system of electrons and ions in a constant
magnetic field, and one finds  a bifurcation  to 
an unstable solution,  at a critical density which is the same as
(\ref{sei}) apart from a factor of order one.

\section{The theoretical  density limit} In order to apply the methods
of statistical mechanics,  dealing with pure Coulomb interactions, we
neglect any boundary
effect, and  just consider  an infinite system of point electrons  in a constant
magnetic field, 
disregarding   the role of the ions. 
If $\vett x_j$ and $\vett v_j$ denote
the position vector and the velocity of the $j$th electron, $e$ its
charge, $\vett B=B\vett e_z$ a constant magnetic field directed along 
the $z$ axis,   and
$\epsilon_0$ the vacuum dielectric constant, the
system of equations of motion in the nonrelativistic
approximation is then 
\begin{equation}\label{uno}
m\ddot{\vect x}_j= e\vect v_j\wedge \vect
B+\frac{e^2}{4\pi\epsilon_0}\sum_{k\neq j}
\frac {\vect x_j-\vect x_k}{|\vect x_j-\vect x_k|^3} \ .
\end{equation} 
It goes without saying, that we are considering here a model describing
an autonomous system, i.e., an isolated one, on which no power is
injected from outside. In terms of fusion devices, this model is presumably 
better suited for high  field devices, which are, in general, 
characterized by large energy confinement
times, and hence need a lower  sustainment.

Obviously, the two forces at the right hand side of (\ref{uno}) play  
opposite roles, the magnetic Lorentz
force  producing ordered motions with confinement, while   the
Coulomb repulsions of the other electrons produce a diffusive
effect,  which we consider as a
perturbation depending parametrically on density. Thus a critical
situation should occur when the Lorentz force and the transverse
component $F_{\bot}$ of the total Coulomb force  somehow balance. 
The size of the Lorentz force, to which only the transverse component
$v_{\bot}$ of the velocity contributes, is immediately estimated by 
$$
v_{\bot}\simeq\sqrt{2k_BT/m}
$$
where $T$ is absolute temperature and $k_B$ the Boltzmann constant.

Much more delicate is the estimate of the vector sum of the Coulomb
forces due to all the other electrons, or rather of the modulus $F_{\bot}$
of its transverse component. This problem can be tackled by statistical methods,
considering  $F_{\bot}$ as a random variable. As $F_{\bot}$  obviously has zero
mean,  according to Chebishev's
theorem  its typical value is given  by its standard deviation
$\sigma_{\bot}$. 
The computations, with respect to the Gibbs
ensemble at a given inverse temperature $\beta=1/k_BT$, 
 will be performed in  the Debye--H\"uckel approximation, i.e, at
 the lowest order in the density.

To compute the standard deviation $\sigma_{\bot}$ of  $F_{\bot}$ (the modulus of the
transverse component of the total Coulomb force acting on a single
electron, the $j$th one), one uses the fact that
the correlations between   ${\vect x_j-\vect x_k}$ and ${\vect
x_j-\vect x_l}$ can be neglected in the Debye--H\"uckel approximation 
(see \cite{imre}), so that  one has
$$
\sigma^2_{\bot}=\frac 23 \left(\frac {e^2}{4\pi\epsilon_0}\right)^2 
\sum_{k\neq j}\,  <\frac 1{|\vect x_j-\vect x_k|^4} >
$$
where  $<\ldots>$ denotes  canonical  average.
On the other hand, the probability  of the relative
distance $r$ between two particles is well known  to be distributed
according to the Gibbs ensemble relative to the effective
Debye--H\"uckel potential
$$
V_{eff}(r)=\frac {e^2}{4\pi\epsilon_0 r} \exp\left(-r/\lambda_D\right)\ ,
$$
where
$$ 
\lambda_D=\sqrt {\epsilon_0 k_BT/(n_e e^2)}\ 
$$ 
is the Debye length.   So eventually at first
order one  finds
\begin{equation}\label{sigma2} 
\sigma_{\bot}^2= \frac 23\frac {e^4}{4\pi\epsilon_0^2}
\frac 1{a^3}\int_0^{+\infty}\frac 1 {r^2}\exp
  \left[-\beta V_{eff}(r)\right] d  r\ .
\end{equation}

Bringing the integrand into dimensionless form one sees  that, if
$a/\lambda_D$ is small, significant values of the integrand  are
assumed only in a region about the  Bjerrum length $b$. 
This is defined (see \cite{meeron2}) by $b=e^2/(4\pi \epsilon_0 k_BT)$,
so that one also has $4\pi b\lambda_D^2=a^3$. So
 the integral in (\ref{sigma2}) is well approximated by the
integral
$$
 \int_0^{+\infty} \frac 1{r^2}  \exp \left(-\frac {b}{r}
 \right)d\, r= \frac {1}{b}=\frac {4 \pi \lambda_D^2}{a^3}\ ,
$$
and  finally one gets 
$$
F_{\bot}\simeq \sigma_{\bot} \simeq \sqrt{\frac23}\,\frac{e^2}{ \epsilon_0}\, \,
\frac {\lambda_D}{a^3}\ .
$$

Thus, in the Debye--H\"uckel approximation   a
balance between confining Lorentz  force and diffusive long range Coulomb
forces on each electron occurs when
\begin{equation}\label{cinque}
B \sqrt{2k_BT/m}\simeq  \sqrt{\frac 23}\,\frac{e}{ \epsilon_0}\,\frac {\lambda_D} {a^3}\ .
\end{equation}
So temperature altogether disappears, and the  transition from order to chaos
 occurs when density and magnetic field are related by
$$
n_e=  3\, \epsilon_0\,\frac {B^2}{ m}\ ,
$$
i.e., by (\ref{sei}), if $\epsilon_0$ is   expressed  in terms of the 
vacuum permeability  $\mu_o=1/(\epsilon_0 c^2)$.

\begin{figure}
  \begin{center}
    \includegraphics[width=\textwidth]{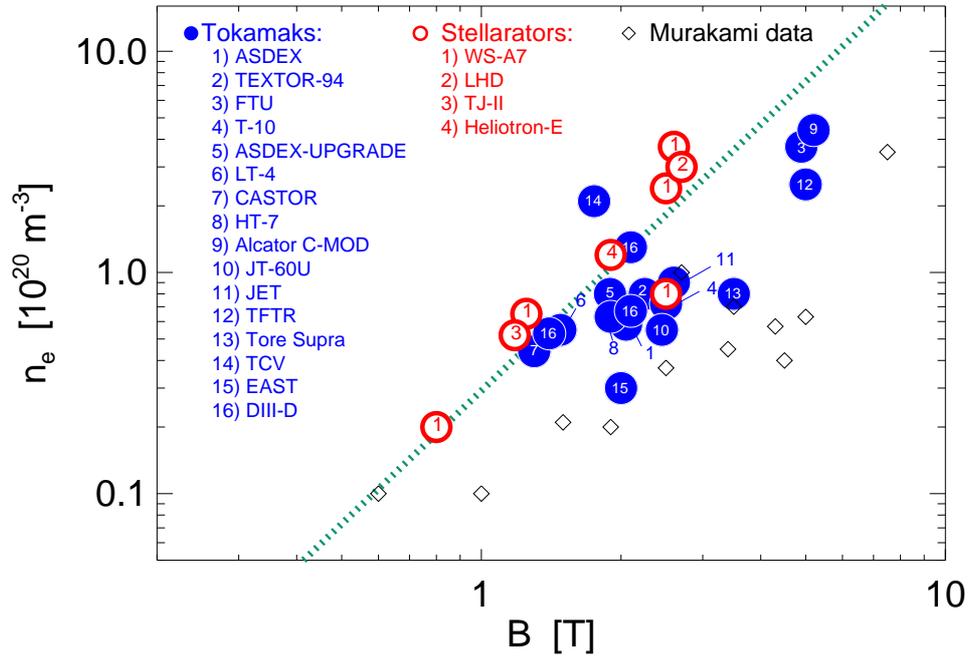}
  \end{center}
  \caption{\label{fig:1} Density limit values \emph{vs}  $B$ for
    various devices:  conventional tokamaks, for which recent data are
    shown (see references from \cite{Stabler} to \cite{Petrie}) along with
    the original ones of Murakami (see \cite{murakami}), 
    stellarator devices (from \cite{Giannone} to
    \cite{Sudo}), and spherical tokamaks (from \cite{Sykes_START} to
    \cite{Kaye_NSTX}). Dotted line is the theoretical   density limit (\ref{sei}). } 
\end{figure}

\section{Comparison with the empirical density limit in fusion
  devices} 

We now  check whether the transition from order to chaos 
discussed here has anything to do with the empirical 
data for collapses in fusion machines.  
A proportionality of the  density limit to the
square of the magnetic field in tokamaks  was  
suggested by Granetz \cite{granetz} on the
basis of empirical data, but apparently was not confirmed by successive
observations \cite{Petrie,greenwald}. 
It is well known that, while at first a proportionality 
to the magnetic field (through $B/R$, where $R$ is the major radius of
the torus)  
had been proposed  on an empirical basis for tokamaks  by
Murakami \cite{murakami}, in the plasma physics community the common 
opinion is rather that
the  density limit for tokamaks should be proportional to the
Greenwald parameter $I_p/r_a^2$, where $I_p$ is the plasma current and
$r_a$ the minor radius of the torus (see \cite{greenwald}). 

We do not enter here a
discussion of this point, and only content ourselves with plotting in 
figure \ref{fig:1}
a collection of available data of  the critical density for several
fusion devices  versus
their operating magnetic field $B$ in log--log scale, comparing the data to the
theoretical formula (\ref{sei}).
One sees that the theoretical law appears to correspond not so
badly to the data  for the high field devices (tokamak and
stellarators), whereas a sensible discrepancy is met for the low
field devices (spherical tokamaks), for which  the 
experimental data are larger by even an order of magnitude.

One should not forget however that we are discussing here a model
describing  an isolated, non sustained, system (i.e., with no input heating
power), whereas  one should expect (see the empirical Sudo limit for
stellarators \cite{Sudo}) that larger densities are accessible as   the 
 input power is increased (although this is not so clear for tokamaks 
\cite{greenwald}). This is illustrated, in the figure, by the
 three points reported for  the same  device (the stellarator 
WS-A7 \cite{Giannone})  at  
 essentially the same applied field, which however  correspond to  three 
different (increasing)  input heatings. 
Now,  the low field devices for which a sensible discrepancy is shown
in the figure,
are just the ones characterized, in general, by lower
confinement time and thus by larger sustainment,  so that  a
discrepancy corresponding to larger experimental values might be expected. It
would thus be of  interest to extend our model by including  some
forcing  describing the operations of sustained devices.

\section{Comments}
In view of the lack of any first principles rationale for  the
existence of a density limit in fusion devices, the partial agreement  
of the theoretical formula (\ref{sei})  with the experimental data seems encouraging. 
In our opinion, a key feature characterizing the present
approach is that, at variance with the treatments 
involving the continuum approximation, such as magnetohydrodynamics, 
we are dealing with the
plasma as a discrete system of 
particles.  Indeed in our treatment a key role is played by  the 
fluctuations of the force acting on
a single particle, and so  the instability found here would be
   lost in the continuum approximation,  or in any
other approximation involving high--frequency cutoffs. 
For an analogous role of discreteness of matter
in cosmology, see \cite{ccg} and \cite{galassie}.

On the other hand, even in plasma physics theory there exists a
literature  in which the discrete nature of matter is taken 
into account. We refer to the works 
(see for example \cite{psimo}, \cite{dubin} and
\cite{rosen})  in which the approach of kinetic theory is followed,
 along the lines of the classical works  of Balescu, Lenard, Lifshitz
 and Pitaevskii. Now, both the   kinetic theoretical approach 
and the statistical mechanical one followed here, are originating from
a common source, namely, the previously mentioned work of Bogolyubov
\cite{bogol}, so that  their results  should agree. Thus the result
found here should in principle be obtained also within the kinetic
theory approach. We leave this interesting problem for 
future possible work.

Finally, it is  worth mentioning that the proportionality of the density limit
to the square of the magnetic field predicted  by the theoretical law (\ref{sei}), if confirmed,
might have relevant implications for future tokamaks.
However, for what concerns the analytical side of the problem,
in order to produce more exact fits with the
experimental data one should  push the  theory much more forward. First
one should settle  the problem of the contribution of the
ions, and include sustainment in the model.
Then one should consider the boundary effects, and the inhomogeneities
of the macroscopic quantities characterizing the plasma, such as
temperature, magnetic field and density.  
More in general, one should consider
relativistic effects, i.e., the retardations of the fields,
and thus emission of radiation.
Furthermore, one should consider
 higher order  perturbation  effects, going 
 beyond the Debye--H\"uckel approximation.
Useful information on
all these problems should also  be obtained through numerical studies,
along the lines for example of the recent work \cite{russi} 
on strongly coupled plasmas.

\vskip 0.7 truecm

\emph{We thank  N. Vianello for fruitful
  discussions. We furthermore thank three  referees, as well as   F. Pegoraro and
  R. Pozzoli,   for stimulating   us, through their criticisms, to undertake a
  statistical mechanical discussion of the problem.}
 
\emph{This work, supported by the European Communities
  under the contract of Association between EURATOM/ENEA, was carried
  out within the framework of the European Fusion Development
  Agreement.}

\end{document}